\documentclass[10pt]{article}

\usepackage{setspace}
\usepackage{amsmath}
\usepackage{amssymb}
\usepackage{amsfonts}
\usepackage{graphicx}
\usepackage[utf8]{inputenc}
\usepackage{slashed}
\usepackage{physics}
\usepackage{mathrsfs}
\usepackage{bbm}	
\usepackage{cancel}	
\usepackage[dvipsnames]{xcolor}
\usepackage{float}      
\usepackage{cite}

\usepackage[margin=1in]{geometry}  
\graphicspath{{figures/}}

\renewcommand{\text}{\textnormal}

\usepackage{tikz}
\usetikzlibrary{arrows,shapes}
\usetikzlibrary{trees}
\usetikzlibrary{snakes}
\usetikzlibrary{matrix,arrows} 													
\usetikzlibrary{positioning}				
\usetikzlibrary{calc,through}				
\usetikzlibrary{decorations.pathreplacing} 
\usepackage[tikz]{bclogo} 					
\usepackage{pgffor}							
\usetikzlibrary{decorations.pathmorphing}	
\usetikzlibrary{decorations.markings}
\usetikzlibrary{intersections}		

\tikzset{
    vector/.style={decorate, decoration={snake}, draw},
    fermion/.style={postaction={decorate},
        decoration={markings,mark=at position .55 with {\arrow{>}}}},
    fermionbar/.style={draw, postaction={decorate},
        decoration={markings,mark=at position .55 with {\arrow{<}}}},
    fermionnoarrow/.style={},
    gluon/.style={decorate,
        decoration={coil,amplitude=4pt, segment length=5pt}},
    scalar/.style={dashed, postaction={decorate},
        decoration={markings,mark=at position .55 with {\arrow{>}}}},
    scalarbar/.style={dashed, postaction={decorate},
        decoration={markings,mark=at position .55 with {\arrow{<}}}},
    scalarnoarrow/.style={dashed,draw},
	vectorscalar/.style={loosely dotted,draw=black, postaction={decorate}},
}

\usepackage[
	colorlinks=true,
	citecolor=green!50!black,
	linkcolor=NavyBlue!75!black,
	urlcolor=green!50!black,
	hypertexnames=false]{hyperref}

\begin{document}
\begin{center}

    {\large \bf On the measurements in Quantum Gravity}

    \vskip .75cm

\text{Juanca Carrasco-Martinez} 
\vspace{0.5cm}

\text{Department of Physics, University of California, Berkeley, California 94720, USA}

\text{Theoretical Physics Group, Lawrence Berkeley National Laboratory, Berkeley, California 94720, USA}

\textit{jc.carrasco@berkeley.edu}

\end{center}

    \vskip .75cm


\begin{abstract}
\noindent 
In this essay, we argue that certain aspects of the measurement require revision in Quantum Gravity. Using entropic arguments, we propose that the number of measurement outcomes and the accuracy (or the range) of the measurement are limited by the entropy of the black hole associated with the observer's scale. This also implies the necessity of modifying the algebra of commutation relationships to ensure a finite representation of observables, changing the Heisenberg Uncertainty Principle in this manner.
\end{abstract}

\vspace{4.0cm}

\begin{center}

\text{Essay written for the Gravity Research Foundation}

\text{2024    Awards for Essays on Gravitation}

\end{center}

\thispagestyle{empty}


\newpage
\pagenumbering{arabic} 

\subsubsection*{Introduction}

In recent years, some aspects of the information paradox formulated by Hawking \cite{Hawking:1974rv, Hawking:1975vcx} have been answered by groundbreaking works \cite{Penington:2019npb, Almheiri:2019psf}. These solutions are based on the holographic principle proposed by G'Hooft and L. Susskind \cite{tHooft:1993dmi, Susskind:1994vu}, which was made concrete by Maldacena in the AdS/CFT correspondence \cite{Maldacena:1997re}. Although this represents a great achievement in understanding black holes, many open questions concerning Quantum Gravity remain, and here we delve into the problem of the observer (or the reference frame), of which some aspects were recently explored in \cite{Witten:2023qsv, Susskind:2023rxm, Witten:2023xze}. The problem can be formulated as follows: In General Relativity, the observer can be arbitrarily lightweight since we expect a negligible back-reaction, allowing for a clear measurement of the system's gravity. However, in Quantum Mechanics the same observer must be sufficiently large to prevent measurement fluctuations. Therefore, these two opposite assumptions of the observer or reference frame need resolution in Quantum Gravity.

In this essay, we argue that the entanglement between an apparatus and the system at the moment of measurement is crucial for addressing this tension in Quantum Gravity. We show that while this entanglement entropy can be infinite in quantum mechanics, once gravitational effects are considered, the holographic bounds must be considered, thus changing important aspects of the measurement. We elaborate on three possible implications: 1) The number of possible outcomes measured by the observer is bounded. 2) The saturation of the bound could correspond to the fact that all quantum apparatus converts into a black hole when they achieve some maximum precision. This also restricts the accuracy of the apparatus ($\Delta$) or the length range ($L$) of the observation, as we prove in a toy model for position measurement. 3) The modification of commutator relationships is necessary since $[\hat{X},\hat{P}]=i\hbar f(\hat{X},\hat{P})$ must have a finite representation and not an infinite representation, assuming that eigenvalues of $\hat{X}$ and $\hat{P}$ operators correspond to possible outcomes. This justifies the Generalized Heisenberg Uncertainty Principle, which was proposed in many models of Quantum Gravity, see \cite{Konishi:1989wk, Maggiore:1993rv, Maggiore:1993kv, Kempf:1994su, Garay:1994en, Fadel:2021hnx}.

\subsubsection*{Measurements in Quantum Gravity}

The modern description of measurement involves a unitary evolution between the apparatus (A) and the system (S), transforming the initial state $\ket{0}_A\ket{\psi}_S$ into the final state $U\ket{0}_A\ket{\psi}_S$. In this context, it is also pertinent to use the Krauss Operators $K_i=\bra{i}_A U\ket{0}_A$ \footnote{Note that $K_i$ is an operator because $U$ acts on both the system and the apparatus Hilbert spaces. As a result, the contraction is confined to the apparatus Hilbert space $\mathcal{H}_A$, leaving $K_i$ as an operator in the system Hilbert space $\mathcal{H}_S$.}, where $U$ is the total evolution of the apparatus and the system. The Positive Operator-Valued Measures (POVMs) are the set of operators defined as $ E_i=K_i^\dagger K_i$, subjected to $\sum_i E_i=1$. Therefore, the final state of the apparatus and the system can be written as follows:
\begin{equation}
U\ket{0}_A\ket{\psi}_S=\sum_i \ket{i}_A K_i\ket{\psi}_S.
\end{equation}
Note that entanglement is an inherent property of this evolution, for instance, the reduced density matrix of the system $\rho_S=\sum_m K_m \ket{\psi}_S\bra{\psi}_S K_m^\dagger$ can be a mixed state starting from a pure state. To illustrate this, let's prepare an electron in a superposition of up and down:
\begin{equation}
\ket{0}_A \frac{1}{\sqrt{2}}(\ket{\uparrow}_S+\ket{\downarrow}_S)\rightarrow \frac{1}{\sqrt{2}}(\ket{0}_A\ket{\uparrow}_S+\ket{1}_A\ket{\downarrow}_S),
\end{equation}
where, in the final state, the apparatus registers 0 (or 1) if it measures the sate $\ket{\uparrow}$ (or $\ket{\downarrow}$). Therefore, the measurement device's reduced state is:
\begin{equation}
\rho_A=\frac{1}{2}\ket{0}\bra{0}+\frac{1}{2}\ket{1}\bra{1}.
\end{equation}
\noindent
As we see, the resultant state of the apparatus typically loses its purity, and it can exhibit a maximum entanglement entropy, which depends on the number of possible outcomes "$\mathcal{N}$" \footnote{Here $\mathcal{N}$is associated to the physical Hilbert space, not the typically larger kinematical Hilbert space. This is because $\mathcal{N}$ is restrained by holographic bounds, which are fundamentally connected to the physical states. Constraints on the dimension of the kinematical Hilbert space are beyond the scope of this paper.} of the apparatus as:
\begin{equation}
S_{max}=\ln \mathcal{N}.
\label{eq2}
\end{equation}
Note that this maximum entropy remains unbounded in Quantum Mechanics since the Bekenstein bound \cite{Bekenstein:1980jp} ($S\leq 2\pi E R/\hbar$) can be adjusted such that the energy $E$ becomes infinite within a finite spherical region of radius $R$.  This is a natural choice for the energy $E$ since we expect no back-reactions of the apparatus when we are measuring, for example, the momentum of an electron. Nevertheless, in the context of quantum gravity, any region with radius $R$ possesses a maximum energy that corresponds to the mass of a black hole, and it gives a holographic bound for the entropy when Bekenstein's conditions are satisfied (see \cite{Bekenstein:1980jp, Bousso:1999xy, Bousso:2002ju}):
\begin{equation}
S_{max}\leq \frac{A}{4G\hbar},
\label{eq3}
\end{equation}
where $A$ is the area that circumscribes the system and denotes its scale.

\begin{figure}[H]
    \centering
    \includegraphics[scale = 0.3705]{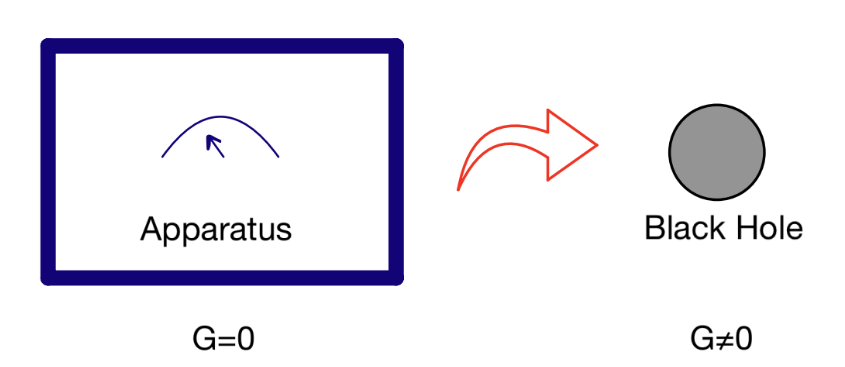}
    \caption{Representation for the possible transition of the ideal apparatus in pure quantum mechanics (where $G = 0$) into a black hole when gravity is turned on ($G\neq 0$) when saturating the number of possible outcomes.}
    \label{fig: Fig 1}
\end{figure}

\noindent
This result is significant as it denotes a possible transition point for the apparatus when it saturates the bound, becoming a hyper-entropic state, so a potential black hole formation must be considered if its area has a negative expansion (see \cite{Bousso:2022cun}), suggesting that black holes and apparatuses could have a correspondence in quantum gravity when we try to get the maximum possible number of results or, in other words when we make the most accurate apparatus (see include \ref{fig: Fig 1}).

Moreover, from \ref{eq2} and \ref{eq3}, we claim that the bound for the number of possible outcomes  is given by:
\begin{equation}
\mathcal{N}\leq \exp[\frac{A}{4G \hbar}].
\label{eq4}
\end{equation}
While this bound reduces when decreasing the gravitational constant ($G$) for a given black hole of mass $M$ ($\mathcal{N}\leq e^{4\pi M^2 G/\hbar}$, where we have used $A=4\pi R^2$ and $R=2MG$), it also reduces with the apparatus's size. Thus, fixing the size of the area $A$, then from \ref{eq4}  as $G \rightarrow 0$ it corresponds to the unbounded situation of quantum mechanics, i.e., $\mathcal{N}\leq \infty$.

\subsubsection*{A toy model for Position Measurement}

Consider the following toy model for position measurement, where the Krauss operators are defined as:
\begin{equation}
K_{x_i}=\int_{0}^{L} \theta(x'-x_i)\ket{x'}\bra{x'}dx',
\label{krauss}
\end{equation}
where $\theta(x)=1 $ if $0\leq x\leq \Delta$, and $\theta(x)=0 $ otherwise.
We assign the possible results $x_i$ such as $$x_1=0,\hspace{0.2cm} x_2=\Delta,\hspace{0.2cm} x_3=2\Delta,\hspace{0.2cm}...\hspace{0.2cm}, x_N=L, $$where $L=\mathcal{N}\Delta$ represents the measurement range.
\noindent
The number of possible outcomes $\mathcal{N}$ must correlate with the precision of the measurement apparatus. For example, for a continuous variable like position measurement, higher $\mathcal{N}$ signifies greater precision, thus, $\Delta$ represents the accuracy of our apparatus.

In this model, the maximal entropy is given by:
\begin{equation}
S=\ln[L/\Delta]\leq \frac{A}{4G\hbar},
\end{equation}
which establishes a fundamental limit on accuracy $\Delta$ in the apparatus due to Quantum Gravity:
\begin{equation}
L e^{-\frac{A}{4G\hbar}}\leq\Delta,
\end{equation}
highlighting the dependence of minimum uncertainty on the apparatus's measurement range $L$ and the entropy $S$ of a black associated with the apparatus. We must point out that this limitation on the precision of the apparatus arises from entropy bounds, indicating that it is a fundamental constraint rather than merely a technological one. Specifically, this limit applies to the Krauss operators in \ref{krauss} and the implicitly defined observable operators. In the next section, we discuss how the commutation operators must be modified by \ref{eq4}, altering the Heisenberg uncertainty principle and therefore the observable operators which could be an explanation for this limit. 

Note that in the absence of gravity, i.e. $G\rightarrow 0$, we make possible a perfect apparatus ($0\leq\Delta$, with $S\rightarrow \infty$ and finite $L$) or an unbounded range ($L \leq \infty$ with $S\rightarrow \infty$ and finite $\Delta$). However, in quantum gravity, there is a limit concerning the accuracy or the size of the system we are measuring, which is dictated by :
\begin{equation}
L\times e^{-\frac{A}{4G\hbar}}\leq \Delta,\hspace{1cm}L\leq \Delta\times e^{\frac{A}{4G\hbar}},
 \label{equation1}
\end{equation}
respectively. The latter limit is reminiscent of the fact that a black hole has a limit for the growth of its interior given by $$L\propto e^{\frac{A}{4G\hbar}},$$ calculated in \cite{Iliesiu:2021ari}. One speculative argument that can potentially connect these two results goes as follows:
"Suppose the most accurate observer, Bob, with $\Delta_{min}$, attempts to measure the interior of a black hole (both with the same entropy $S$). Then, suppose Bob defines the interior size of a black hole using the maximum extension $L$ of a given particle within it. In that case, he will eventually reach the limit of \ref{equation1} as the maximum size of the interior for its measurement. Thus, without the possibility of another observer measuring a larger size, the maximum observable length of the interior will be $L\propto e^{S}$, which corresponds to the actual maximum size calculated in \cite{Iliesiu:2021ari}".

Additionally, it is important to consider that this uncertainty extends to measurements of discrete variables such as the polarization of light (vertical and horizontal) or spin $1/2$ measurements, which appear to have only two possible outcomes. However, in the case of polarization measurements, for instance, achieving precisely two results requires the precise alignment of a polarizer in a specific direction. Yet, this assumption isn't entirely accurate, as this alignment is linked to a continuous position variable, introducing uncertainty and creating more possible outcomes. Therefore, the limit on the accuracy of the position measurement can potentially propagate to all kinds of measurements. Also, we must point out that in this example, it becomes evident that the mass of the polarizer is important to decrease the uncertainty, as the photon's momentum has a lower impact on it if the polarizer were heavier.

\subsubsection*{Modifications on  Commutator Relationships}

Another interesting implication from the limit \ref{eq4} of the number of possible outcomes $\mathcal{N}$  is the necessity to adjust the commutator relations if we still assume that the eigenvalues of $\hat{X}$ and $\hat{P}$ give the outcomes of the measurements since the conventional relationship $[\hat{X},\hat{P}]=i\hbar$ does not admit a finite representation. In general, we must impose:
\begin{equation}
[\hat{X},\hat{P}]=i\hbar f(\hat{X},\hat{P})\approx i\hbar (1-\alpha g(\hat{X},\hat{P})),
\label{eq5}
\end{equation}
where $f$ is a function with an $\mathcal{N}$-finite representation, which could be expanded around its quantum mechanics limit ($[\hat{X},\hat{P}]=i\hbar $) with $\alpha=0$, i.e. $f(\hat{X},\hat{P})_{\alpha=0}=1$. Then, taking $\mathcal{N}=e^{S}$  as the number of possible outcomes, we obtain the conditions for these functions by taking the trace of both sides of equation \ref{eq5} in any basis of the $\mathcal{N}$-dimension representation:
\begin{equation}
\mathrm{Tr}([\hat{X},\hat{P}])=i\hbar \mathrm{Tr}[f(\hat{X},\hat{P})]=0\rightarrow \mathrm{Tr}[g(\hat{X},\hat{P})] =\frac{e^{S}}{\alpha}.
\end{equation}

It is intriguing to point out that similar adjustments to the commutation relationship have been proposed, inspired by the Generalized Heisenberg Principle, which represents an extension of the Heisenberg Uncertainty Principle when the effects of gravity are considered:
\begin{equation}
\Delta x \Delta p\geq \frac{\hbar}{2}+\beta \Delta p^2,
\end{equation}
where $\beta$ denotes a constant that depends on the specific string theory or loop quantum gravity model. For further details, refer to \cite{Konishi:1989wk, Maggiore:1993rv, Maggiore:1993kv, Kempf:1994su, Garay:1994en, Fadel:2021hnx}. 

Finally, although this result may seem unrelated to the apparatus accuracy limit, as we have noted, both modifications impact the level of observable operators. Then, a follow-up paper will present a more formal exploration of the connection between these two consequences of the bound \ref{eq4} on the number of outcomes.

\subsubsection*{Conclusion}

In conclusion, the exploration of Quantum Gravity through entropy bounds reveals profound implications for measurement processes. We proved that the number of possible outcomes is bounded, that apparatuses possess a given maximum precision in Quantum Gravity, and also suggested a possible deep connection between measurements and the formation of black holes. Finally, it is also fascinating to note that, inspired by this, we may be able to establish conceivable connections between some black hole physics results and corrections to the Heisenberg Uncertainty Principle.

\subsubsection*{Acknowledgements}
We thank Oscar Sumari, Daniel Carney, Luca V. Iliesiu, and Geoff Penington for their useful comments.

\medskip

\bibliographystyle{unsrt}
\bibliography{mibibliografia}

\end{document}